\documentstyle[12pt,psfig]{ioplppt}          % use this for preprint style
\def\beq{\begin{equation}}
\def\eeq{\end{equation}}
\def\ds{\displaystyle}
\def\beqar{\begin{eqnarray}}
\def\eeqar{\end{eqnarray}}
\def\B{\rm B}

\def\s{\rm s}
\begin{document}
\title{
Local Thermal and Chemical Equilibration and the Equation of State 
in Relativistic Heavy Ion Collisions}[Equilibration and EOS in HIC's]
\author{L V Bravina\dag\ftnote{6}{Alexander von Humboldt Foundation 
Fellow}\ftnote{7}{Permanent address: Institute for 
Nuclear Physics, Moscow State University, Russia},
M Brandstetter\dag, 
M I Gorenstein\dag\ddag\ 
\ftnote{9}{Permanent address: Bogolyubov 
Institute for Theoretical Physics, Kiev, Ukraine},\\
E E Zabrodin\dag*, M Belkacem\dag$^+$,
M Bleicher\dag, 
S A Bass\S\ftnote{8}{Feodor Lynen Fellow of the Alexander von 
Humboldt Foundation}, C. Ernst, 
M Hofmann\dag, 
S Soff\dag$\Vert$, 
%C Spieles\P$\sharp$,
%H Weber\dag, 
%UrQMD group, 
H St{\"o}cker\dag, and W Greiner\dag }
\address{\dag\
Institut f\"ur Theoretische Physik, Goethe Universit\"at
    Frankfurt, Germany}
\address{\ddag\
School of Physics and Astronomy, Tel Aviv University, 
    Tel Aviv, Israel}
\address{\S\
Dept. of Physics, Duke University, Durham, USA}
\address{$\Vert$\
Gesellschaft f\"ur Schwerionenforschung, Darmstadt, Germany}
%\address{\P\
%Nuclear Science Division, LBNL, Berkeley, CA 94720, USA}

\begin{abstract}
Thermodynamical variables and their time evolution
are studied for central relativistic heavy ion collisions from 10.7 
to 160 AGeV in the microscopic Ultrarelativistic Quantum 
Molecular Dynamics model (UrQMD).
The UrQMD model exhibits drastic deviations from equilibrium during 
the early high density phase of the collision. Local thermal and chemical 
equilibration of the hadronic matter seems to be established only 
at later stages of the quasi-isentropic expansion in the 
central reaction cell with volume 125 fm$^{3}$. Baryon energy 
spectra in this cell are reproduced by Boltzmann distributions at 
all collision energies for $t\geq$10~fm/$c$ with a unique rapidly dropping 
temperature. At these times  the equation of state has a simple form:  
$P \cong (0.12-0.15)\, \varepsilon$.  At SPS energies
the strong deviation from chemical equilibrium is found for mesons,
especially for pions,  even at the late stage of the reaction.
The final enhancement of pions is supported by experimental data.
\end{abstract}

  %journal style
\maketitle                               %journal style

\section{Introduction}
The main goal of relativistic heavy ion experiments at Brookhaven 
and CERN is to study the properties of strongly interacting hot and 
dense hadronic matter.
The description of nuclear (as well as hadronic) collisions may be
achieved in terms of semi-phe\-no\-me\-no\-logical models, 
which may be subdivided into macroscopic and microscopic models.
Macroscopic models, like well-known hydrodynamic 
\cite{Landau,Twofl,BMGR,Strott,hydro} or thermal 
\cite{Fermi,Hagedorn,Braun,Goren} 
models use a few thermodynamic quantities like energy density,
temperature, pressure, chemical potential,
etc., to describe the macroscopic properties of a system of
colliding nuclei. But the applicability of thermodynamics is 
based heavily on the hypothesis of local (or rather global)
thermodynamical equilibrium (LTE) in the system. Therefore, all 
macroscopic models have to adopt the LTE as an {\it ad hoc\/} 
assumption which has not been proved yet.
The question remains still open despite the intensive
investigations 
(e.g. \cite{Landau,Shur72,Hofmann,Geiger,Heinz}
and references therein).
So, is the agreement between the predictions of macroscopic models 
and experimental data accidental or not?

In contrast to the macroscopic approaches, the various Monte-Carlo
microscopic string-, transport-, cascade-, etc. models  
\cite{urqmd1,QGSM,RQMD,ARC,ART} are based only on the assumptions
made about the interaction mechanisms between the constituents,
particles or partons. These models do not apply the LTE and sharp
freeze-out hypotheses 
or any macroscopic characteristics of the system. Therefore, it
looks very promising to check the approach to the LTE and to extract 
the Equation of State (EOS) at different stages of relativistic 
heavy ion reactions by means of one of the microscopic models. Here, 
we employ for  such analysis the newly developed Ultrarelativistic 
Quantum Molecular Dynamics (UrQMD) model \cite{urqmd1}. 

\section{Ultrarelativistic Quantum Molecular Dynamics}
UrQMD is a many-body transport model designed to describe heavy ion 
collisions in the laboratory energy range from several hundreds of MeV 
to several TeV per nucleon. 
The model treats binary elastic and inelastic hadronic collisions 
as well as many-body resonance decays.
Both processes drive the system  towards thermodynamical equilibrium. 
However, the time scales may be too short in heavy ion collisions 
for actually achieving LTE.

The UrQMD model is based mainly on the concept of string and resonance 
excitation. At low and intermediate energies the description of $hh$
or $AA$ collisions may be achieved in terms of interactions between
the particles (hadrons) and their excited states (resonances). 
There are 55 baryon and 32 meson states as discrete degrees of freedom 
in the model as well as their antiparticles and explicit 
isospin-projected states with masses up to 2.25 GeV/$c^2$. 
The experimental hadron cross sections and resonance decay widths are 
used when available.
At high energies the picture of excited strings stretched between the 
constituents is appropriate for understanding the processes of 
multiparticle production. Hadrons, produced through the string decays, 
have a non-zero formation time which depends on their four-momentum.
Newly produced particles cannot interact during their formation time
except the leading hadrons, which are allowed to
interact with the reduced cross sections, proportional to the number 
of their original valence quarks.  
The Pauli principle is applied to hadronic collisions by blocking the 
final state if the outgoing phase space is occupied.
No Bose effects for mesons are implemented in the present
version of UrQMD. 
The detailed presentation of UrQMD together with a comparison with 
experimental data is given in reference~\cite{urqmd1}. 

As well as other transport microscopic models, UrQMD was tuned to 
describe properly the main features of hadronic collisions. Then
the model for {\it hh} collisions was put into the dynamical 
transport scheme. Some additional parameters like formation time for 
non-leading particles appeared. In its present version %(without the
%mechanism of strangeness enhancement), 
UrQMD reproduces nicely the
spectra and average characteristics of nucleons and pions, but 
fails to describe the production of strange baryons, antibaryons
and anti-kaons \cite{urqmd1}.

\section{General Characteristics of A+A Collisions}
\begin{figure}  %Freeze-out of different particles.
\psfig{file=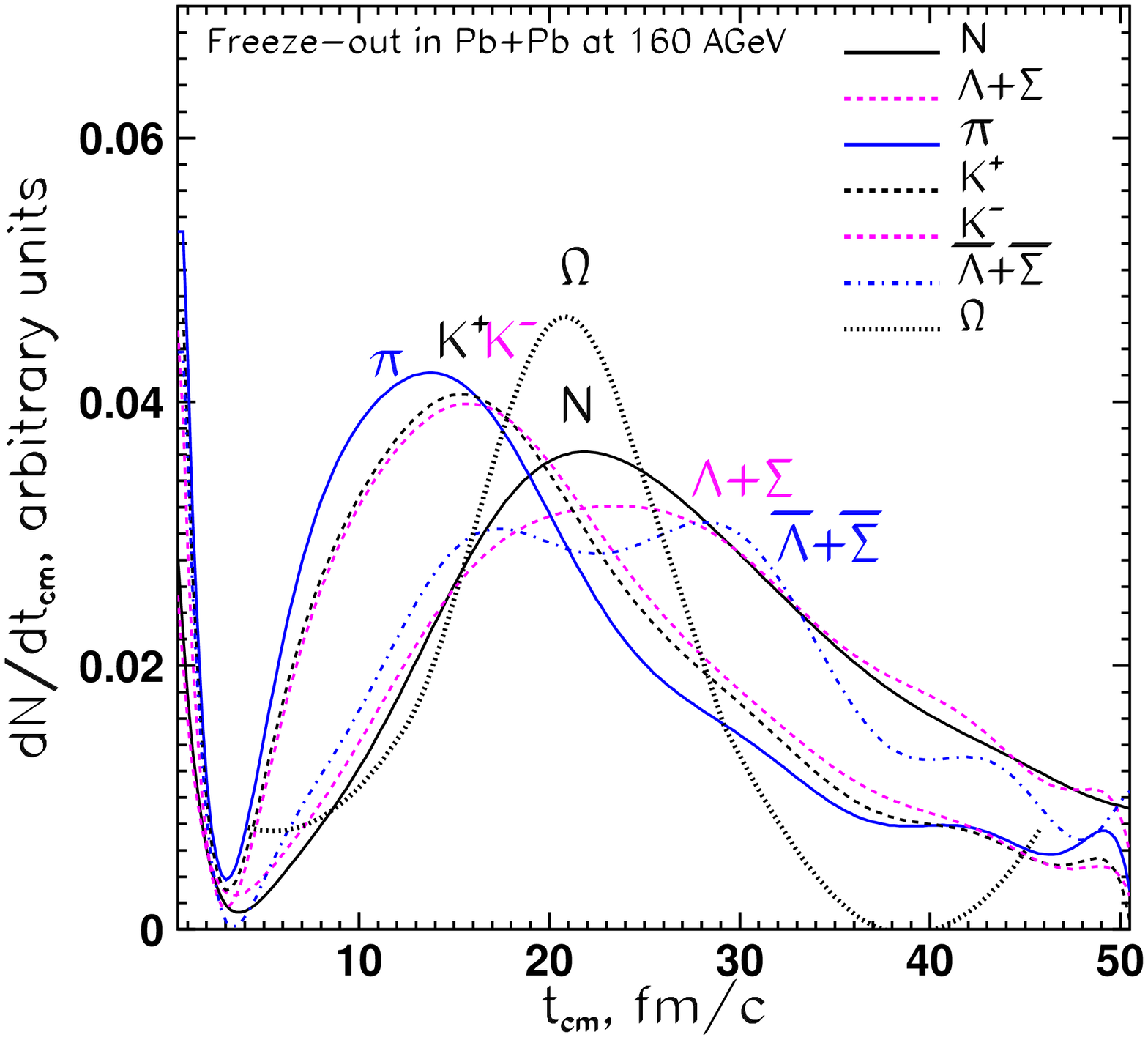,width=16cm}
\caption{ 
$dN/dt_{\rm cm}$ distribution of the final state hadrons
over their last interacting points produced in central (b=0)
Pb+Pb collisions at 160 AGeV. The distributions are normalized
to unity. The baryons
($N$, $\Lambda + \Sigma$, $\bar{\Lambda}+\bar{\Sigma}$, $\Omega$)
are emitted later then mesons ($\pi$, $K$, $\bar K$),
according to their hadronic interaction cross sections.
}
\label{fig1}
\end{figure}

%\subsection{Freeze-out of particles} 
{\bf Freeze-out of particles.} 
Usually, all macroscopic models use the Equation of State (EOS), i.e.
$P = P(\varepsilon, \rho_{\B})$, where $P$ is the pressure, 
$\varepsilon$ the energy density and $\rho_{\B}$ the baryonic
density, as a parametrization \cite{Greiner}, and
then apply an instantaneous freeze-out of hadrons, produced from many
different space-time cells.   The freeze-out conditions for these
cells (temperatures, baryonic chemical potentials and collective
velocities) are  however model dependent and may be rather different (see e.g. 
\cite{bravina95,Shur95,Matiello9597,Bass1}). 
From figure~\ref{fig1}, which displays the $\d N / \d t_{cm}$ distributions
of the emitting times of different hadronic species in Pb+Pb 
collisions at 160 AGeV, normalized to unity, one may conclude that 
the average `life-time' of the hadron system is approximately
25 fm/$c$. The maximum of the meson freeze-out distribution is earlier than that
for baryons. 
Due to the high meson multiplicity, nevertheless, the number of mesons 
freezing out at the later reaction stages exceeds that of baryons. 
%Mesons %freeze out earlier than baryons. 
Apparently, there is
no sharp freeze-out in the system. The time delay in the baryon 
freeze-out is controlled by the cross-sections for meson-baryon  
interactions. The smaller the cross-section, the earlier 
the freeze-out, and vice versa.  This can be easily seen 
when comparing the freeze-out distribution of the $\Omega$  with that 
of the nucleons, since the $\Omega \pi$ cross-section
is only the half of the $N \pi$ one.                           

\begin{figure}  %Evolution of strangeness
\psfig{file=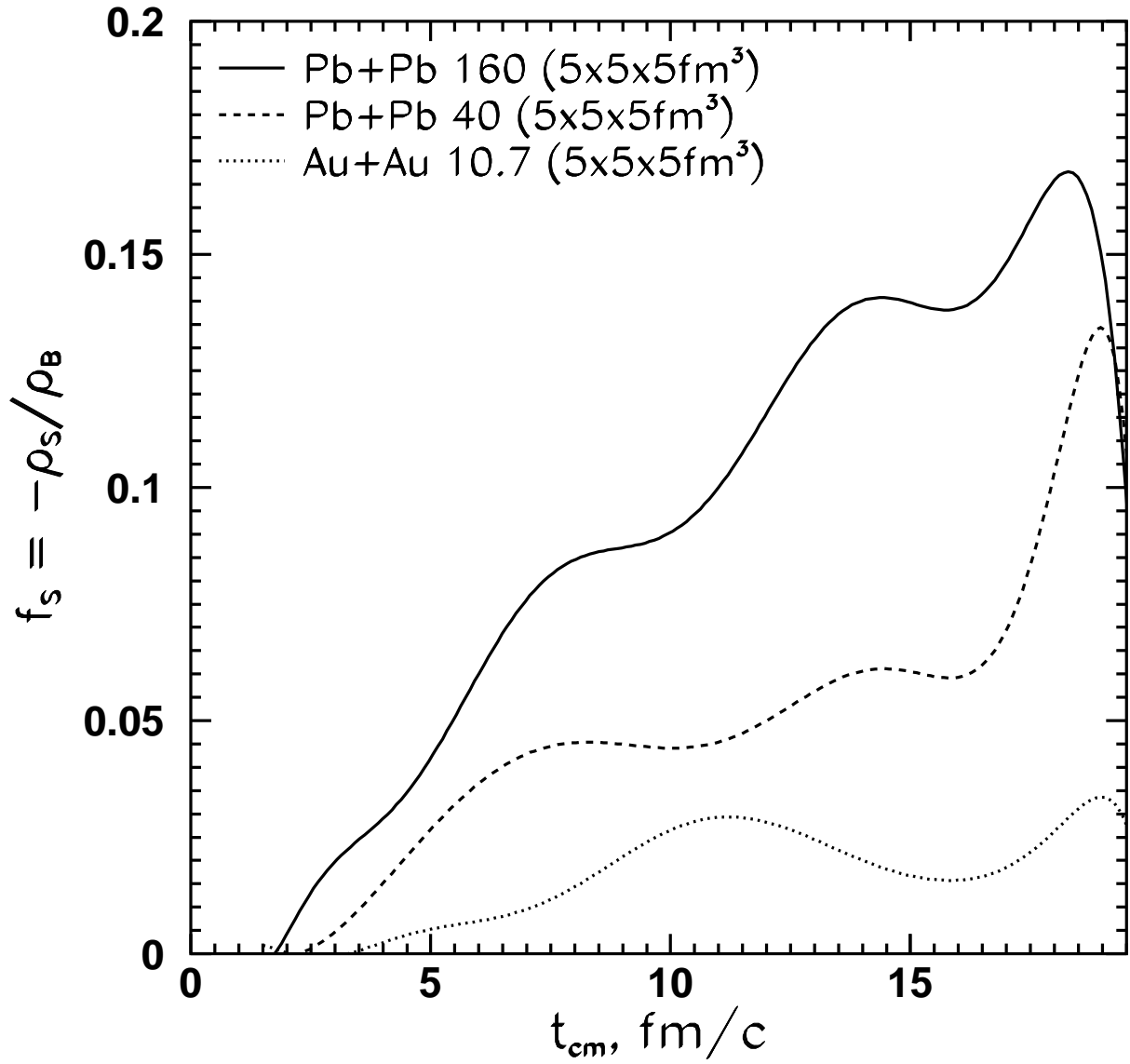,width=16cm}
\caption{ 
Time evolution of strangeness per baryon $f_{\s}=-\rho_{\rm S} /
\rho_{\B}$, obtained in central cell of volume $V=5\times5\times5$ 
fm$^3$, in A+A reactions at 10.7, 40 and 160 AGeV.
}
\label{fig3}
\end{figure}

%\subsection{Strangeness production} 
{\bf Strangeness production.} 
In the central region of heavy ion collisions  
the evolution behaviour of the strangeness density 
$\ds \rho_{\rm S}=\sum_{i} S_i\cdot n_i$, where $S_i$ and 
$n_i$ are the strangeness and density of the hadron species $i$, 
can not be explained by simple combination of $K$, $\bar{K}$ and $\Lambda$, but
is defined by contributions of all species, carrying the strange charge. 
At 10.7--160 AGeV the total strangeness density of all objects in the
central zone 
is small and negative. This result is independent on the size of 
the cell. This is due to the fact, that kaons which are mainly produced 
together with lambdas escape from the interaction zone much earlier than $\Lambda$'s
and $\bar{K}$ because of their small interaction cross section with hadrons.
In the small cells the strangeness density has minimum 
somewhere at the maximum overlap of the nuclei and then it relaxes 
to zero. For the ratio $f_{\s}=-\rho_{\rm S}/\rho_{\B}$ 
(figure~\ref{fig3}) the behaviour 
is different. This ratio rises continuously with time because the
baryon density decreases much faster than the strangeness density. 

At the early stages of the reaction the strange charge is carried 
mostly by resonances. In contrast to AGS energy reactions, 
at SPS energy the contribution of strange baryons to the total
strangeness is relatively small, so the strange charge in the cell 
is defined mostly by meson contributions.

\section{Statistical Model of Hadron Ideal Gas}
To figure out whether one can reach the deconfinement phase
at the particular collision energy one should know the densities
and temperatures which the system reaches during the reaction.
In our scheme the thermodynamical parameters of the system  -- 
temperature $T$, baryonic chemical potential  
$\mu_{\B}$ and strange chemical potential $\mu_{\rm S}$ --
were extracted at each time-step of the UrQMD evolution. 
%of the colliding system were extracted from 
%the predictions of 
We use the Statistical Model of Ideal Gas (SM) with the same
55 baryon and 32 meson species and their antistates 
considered in UrQMD model. As an 
input for the SM,  the  energy density $\varepsilon$, baryon 
density $\rho_{\B}$ and strangeness density $\rho_{\rm S}$ are defined  
by the UrQMD for the central cell 
%with volume $V$ 
in A+A system:
\beqar
\ds
\varepsilon^ &=& \sum_i 
            \varepsilon_i(T,\mu_{\B},\mu_{\rm S}) \\ 
\rho_{\B}^&=& \sum_i 
            B_i\cdot n_i(T,\mu_{\B},\mu_{\rm S}) \\
\rho_{\rm S}&=& \sum_i 
            S_i\cdot n_i(T,\mu_{\B},\mu_{\rm S})
\quad.
\label{smigeq}
\eeqar
Here  $B_i$, $S_i$  
are the baryon charge and strangeness of the hadron species $i$,  whose
particle  densities,  $n_i$,  and 
energy densities,  $\varepsilon_i$,   
are calculated  
within the 
Statistical Model as:
\beqar
\ds
n_i
 = \frac{d_i}{2\pi^2\hbar^3}\int_0^{\infty}p^2f(p,m_i)\d p \\
\varepsilon_i
 = \frac{d_i}{2\pi^2\hbar^3}\int_0^{\infty}
p^2 \, \sqrt{p^2+m_i^2}\, f(p,m_i)\d p \\
f(p,m_i) =  \left[ 
\exp\left(\frac{\sqrt{p^2+m_i^2}-\mu_{\B} B_i-\mu_{\rm S} S_i}{T}\right) +\eta 
\right]^{-1}\quad,
\label{smig}
\eeqar
where $p$, $m_i$  and $d_i$ are the momentum, mass and the degeneracy factor
of the hadron species $i$. $\eta=+1(-1)$ stands for fermions (bosons). Then,
instead of Fermi-Dirac or Bose-Einstein distributions we use for all 
hadronic species the classical Boltzmann distribution $f(p,m_i)$ with 
$\eta=0$. At temperatures above 100 MeV the only visible difference 
(about 10\%) between quantum and classical descriptions is in the 
yields of pions.

The hadron pressure is given in SM 
%the statistical model 
by
\begin{equation}
P(T,\mu_B,\mu_S)~=~\sum_i \frac{d_i}{2\pi ^2\hbar^3}\int_0^{\infty}
p^2 dp~\frac{p^2}{3(p^2+m_i^2)^{1/2}}~f(p,m_i)~.
\label{pressure_sm}
\end{equation}
The entropy density $s=s(T,\mu_B,\mu_S)$ can  be calculated 
%the ideal gas model
%using the microscopic $\varepsilon^{\rm mic}$, $\rho_B^{\rm mic}$, 
%$\rho_S^{\rm mic}$ as input,
%by 
from the thermodynamical identity: $s=(\varepsilon + P -\mu_B\rho_B - \mu_S\rho_S)/T$.
%\begin{equation}
%\varepsilon^{\rm mic}=
%T^{\rm SM}s^{\rm SM}+\mu_B^{\rm SM}\rho_B^{\rm mic}+\mu_S^{\rm SM}\rho_S^{\rm mic}-P^{\rm SM}. 
%\label{entropy_sm}
%\end{equation}

\section{Thermal and Chemical Equilibrium}  
Now we turn to study to what extent the LTE can be 
reached at least in a small cell around the origin of central 
($b=0$ fm) A+A collisions at 10.7, 40 and 160 AGeV. 
To extract the EOS we use the scheme described in previous section similar to 
one, developed in \cite{bravina98} for Au+Au collisions at 10.7 AGeV. Here we 
present new results obtained mainly for SPS energies.

The size of the cell at a certain c.m.s. time was chosen to have 
minimum gradient of the flow velocities in the cell. 
%In Pb+Pb 
%collisions at SPS, for instance, for the cell with $0 \le z \le 1$ fm 
%the average nucleon velocity is only $v_{\rm fl}$=0.25$c$ already at 
%t=3 fm/$c$. This corresponds to the
%temperature $T_{\rm fl}\approx m_{\N}\cdot v_{\rm fl}^2/3=20$ MeV.
The global characteristics of the cell  --  
$\varepsilon$, $\rho_{\B}$,   
$\rho_{\rm S}$ -- are obtained from dynamical UrQMD calculations 
at different time steps of the reaction.
They are put in the left-hand sides of Eqs.~(1-3) 
to find the three unknown variables: $T$, $\mu_B$  and
$\mu_S$.

\begin{figure}  %EOS $E_tot/A(\rho_B)$
\psfig{file=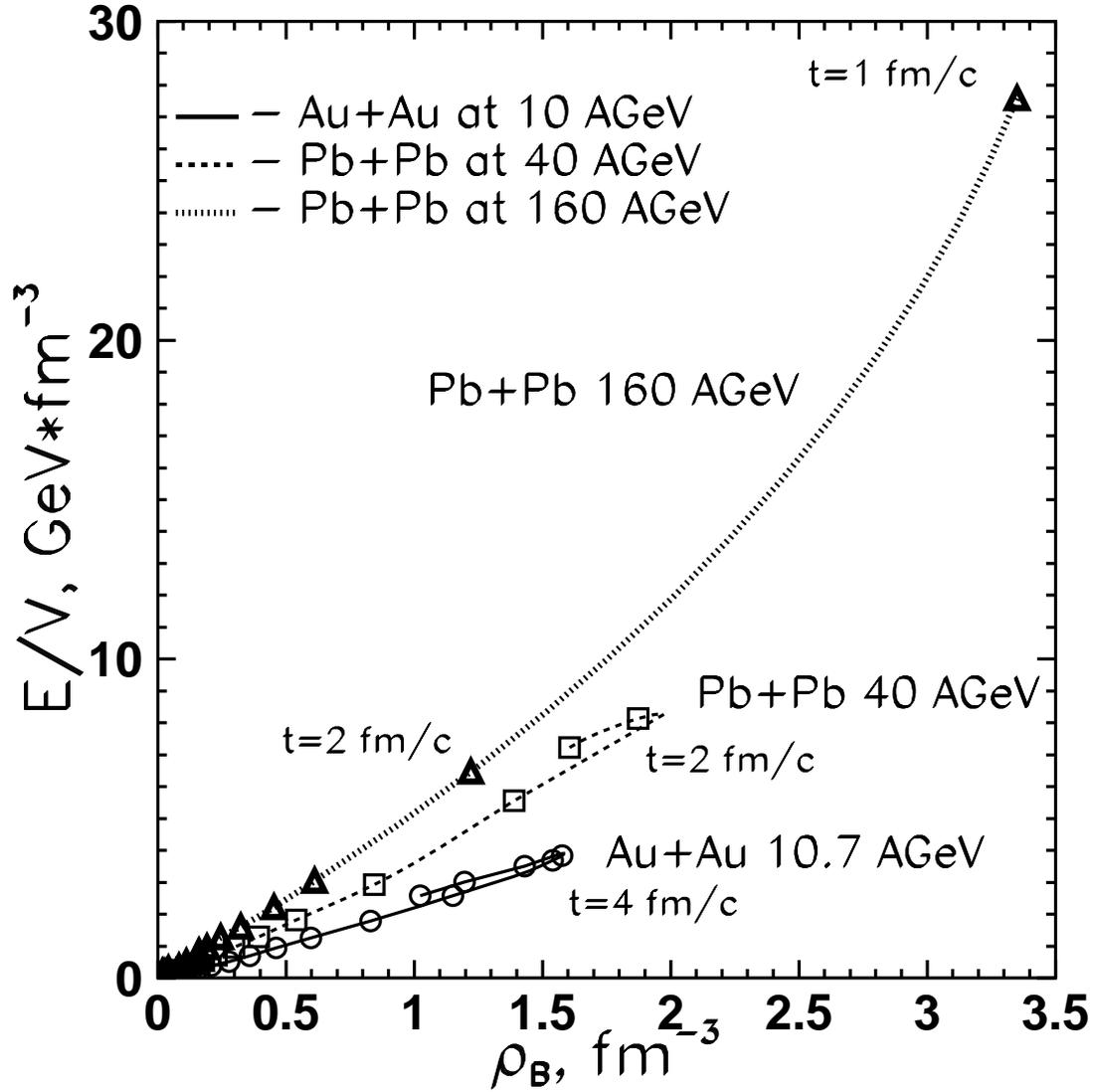,width=16cm}
\caption{
The total energy per volume $E/V$ versus baryon density $\rho_{\B}$,
obtained in the central cell during the time evolution of central
(b=0) A+A collisions at 10.7 (V=125 fm$^3$), 40 (V=32 fm$^3$) and 160 AGeV
(V=16 fm$^3$).
Time $t<10$ fm/c corresponds to the non-equilibrium stage of the reactions.
}
\label{fig5}
\end{figure}

Figure~\ref{fig5} shows the time evolution of  the total energy per 
volume, $E/V$, (including freely streaming particles), 
versus baryonic density $\rho_{\B}$ in 
the central zone of A+A collisions. 
On the non-equilibrium stage of the reaction 
the local energy per volume, $E/V$, strongly depends on
the volume of the cell, V,  and at the beginning of Pb+Pb reaction at 160 AGeV
reaches up to $E/V=27$ GeV/fm$^3$. 
It is much larger than average energy density obtained for  the same reaction in 
\cite{weber98} which at mid rapidity is not more than 4 GeV/fm$^3$. 
The total energy per volume strongly increases with collision energy.
At the late stages at $t=10-20$ fm/c when we expect the equilibrium,
for all reactions the energy density $\varepsilon=E/V$ 
is about 0.6 GeV/fm$^3$.
%One can see the clear similarity 
%between different reactions: UrQMD predicts that within a large time 
%interval ratio $E_{\rm tot}/A$ weakly depends on time.  
%At SPS (160 AGeV) 
%The  energy per baryon in this regime is about 5.5 GeV, at 40 AGeV 
%it stays about 3.5 GeV, while at AGS (10.7 AGeV) it 
%reaches 2.5 GeV only. 
%$\epsilon$
%while the baryon density in the central zone of the 
%collision is not larger then 1.5 fm$^{-3}$ for all three reactions 
%when the energy per baryon is constant. 
It is worth to note that
at AGS energies we deal with baryon rich matter, where about 70\% of 
the total energy is carried by baryons, while at SPS most of the 
energy is deposited in the mesonic sector (more then 70\%). At 
40 AGeV the mesonic and baryonic parts are equal.

\begin{figure}
\psfig{file=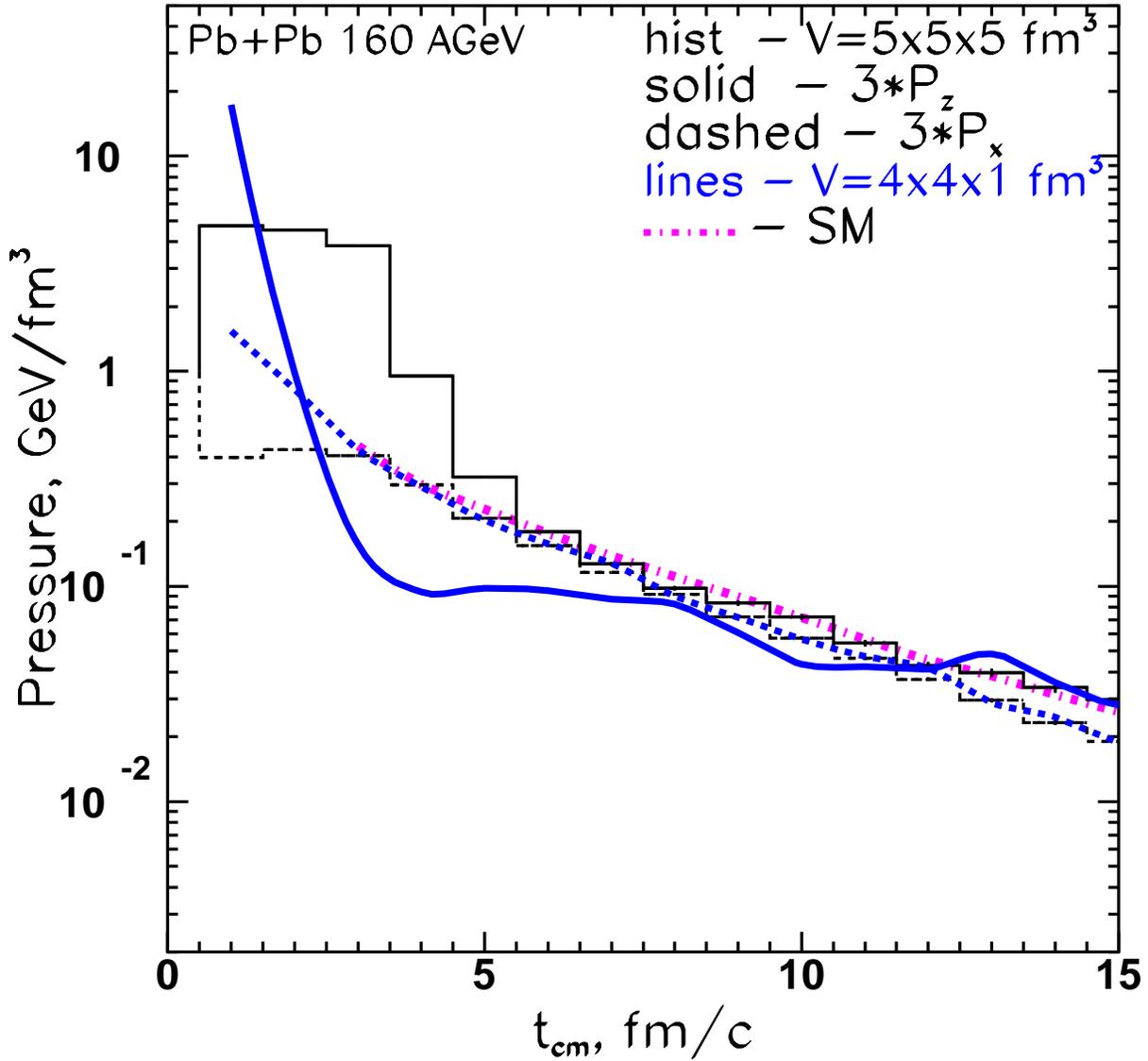,width=16cm}
\caption{ 
The longitudinal ($3\cdot P_{\{z\}}$, solid histogram) and the 
transverse ($3\cdot P_{\{x,y\}}$, dashed curves) diagonal 
components of the microscopic pressure tensor (Eq.~\ref{pressure_micr})  
in the central cell,
$\!5\!\times \!5\!\times \!5$~fm$^3$  (histograms) 
and $4\!\times \!4\!\times \!1$~fm$^3$ (lines), of Pb+Pb collisions 
at 160 AGeV calculated from the virial theorem are compared to
the ideal gas pressure (Eq.~\ref{pressure_sm}) (bold dash-dotted line).
For both cells the equilibration of the whole hadronic system
in the cell takes place only at 8-10 fm/$c$ when pions and other 
mesons become equilibrated. SM nicely reproduces the microscopic
calculations for $5\times5\times5$ fm$^3$ cell from $t>6$ fm/c.
}
\label{fig6}
\end{figure}

\begin{figure}  %Isotropy checking
\psfig{file=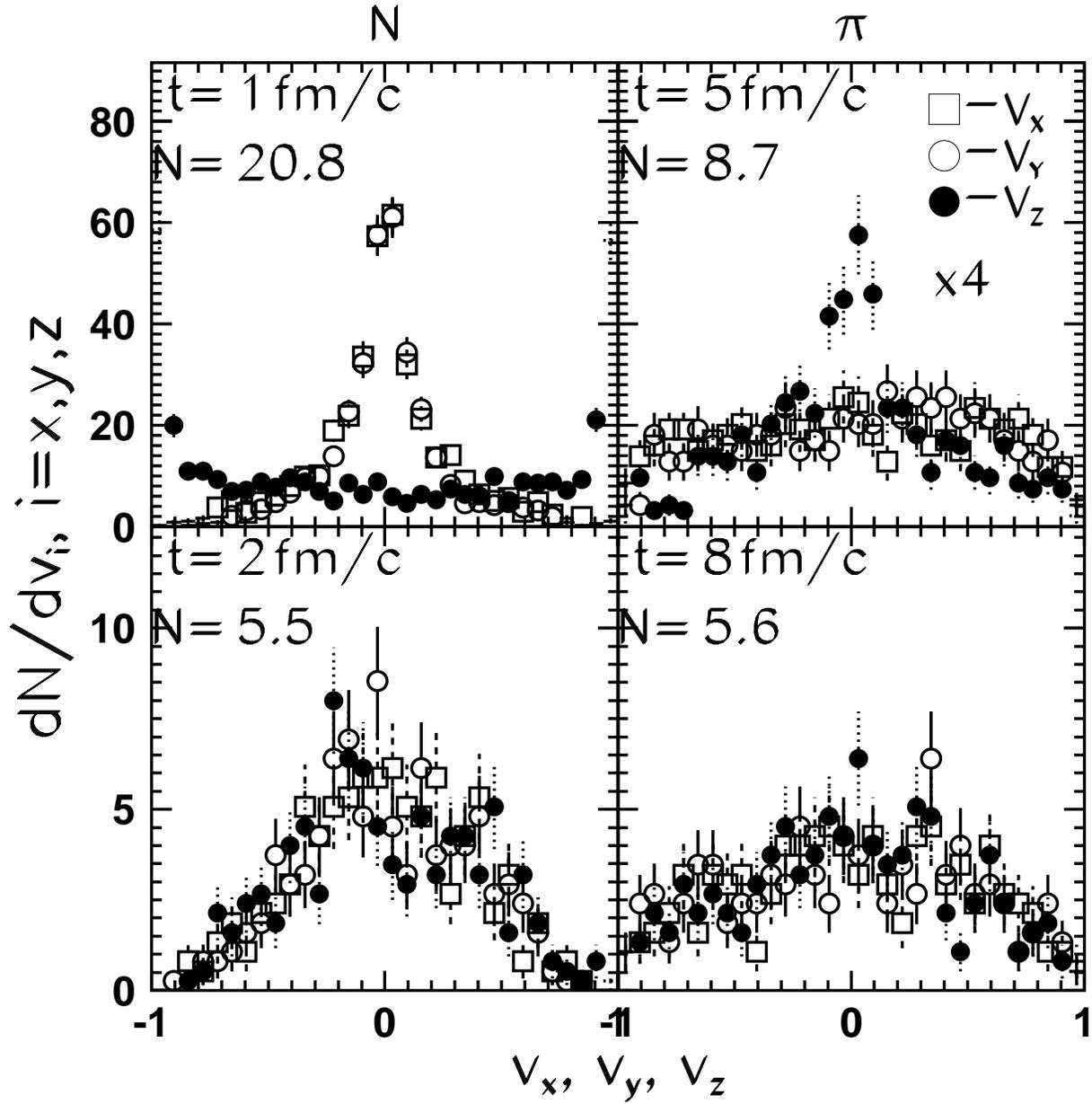,width=16cm}
\caption{ 
Nucleon (left frame) and pion (right frame) velocity
distributions
$dN/dv_i$
($i=z$ ($\bullet $), $x$ ($\Box$) and $y$ ($\bigcirc$))
in central cell of volume $4\times 4\times 1$ fm$^3$ in Pb+Pb 
collisions at 160 AGeV
at $t=$1(5) (upper frame) and  2(8)  (lower frame) fm/$c$.
}
\label{fig7}
\end{figure}

It should be by no means clear that LTE cannot be reached in A+A collisions 
earlier than the certain time 
$t^{\rm cross} = 2 R /(\gamma_{\rm cm} v_{\rm cm})$ 
during which the 
freely streaming nuclei would have passed through each other. Then,
thermalization of the hadron  matter in the cell would 
manifest in the isotropy of pressure, 
\begin{equation}
P^{\rm mic}_z = P^{\rm mic}_y =
P^{\rm mic}_x
=\sum_h
\frac{p^2_{h\{x,y,z\}}}{3V(m_h^2~+~p^2_h)^{1/2}}~,
\label{pressure_micr}
\end{equation}
(where V is volume of the cell,
$p_h$ represents the particle momentum and the
sum in Eq.~(\ref{pressure_micr}) is taken
over all hadrons $h$ \cite{berenguer,sorge})
and more clearly in the velocity distributions of hadrons. 
Figure~\ref{fig6}
depicts the time evolution of pressure $P^{\rm mic}$ 
in longitudinal and transverse
directions together with the predictions of SM for Pb+Pb collisions at
SPS (Eq.~\ref{pressure_sm}). 
The difference between $P_{\perp}^{\rm mic}$ and $P_{\parallel}^{\rm mic}$ 
totally vanishes after  $t \cong 8$ fm/$c$. At this time the microscopic pressure 
%even in small 
in the cell becomes nearly isotropic and approximately 
equal to the ideal gas one. The longitudinal and transverse velocity 
distributions of nucleons in a small cell ($4\times 4\times 1$ fm$^3$)
seem to be equilibrated already at $t = 2$ fm/$c$ (figure~\ref{fig7}).
In contrast, pions are equilibrated at $t \cong 8$ fm/$c$, 
irrespective to the size of the cell. The origin of such a delay is a 
large formation time for non-leading particles. At 
$t = 10$ fm/$c$ the newly produced particles in the cell have undergone
from 2 to 6 elastic collisions, while the baryons have suffered
more than 15 interactions.

\begin{figure}  %Energy spectra in UrQMD cell and in SM
\psfig{file=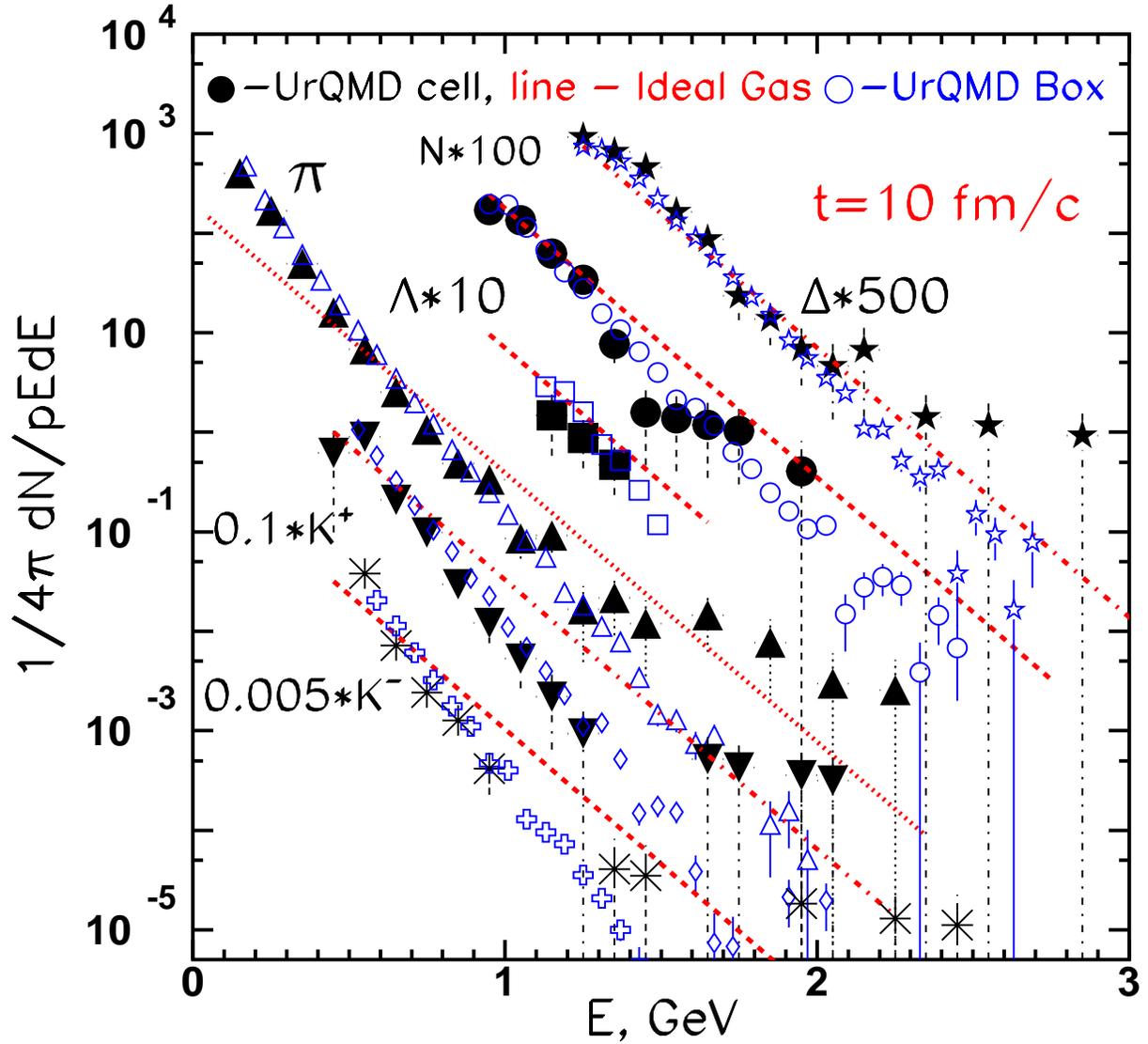,width=16cm}
\caption{ 
Energy spectra of $N$ ($\bullet $), $\Lambda $ ($\Box $),
$\pi $ ($\bigtriangleup $), $K^+$ ($\bigtriangledown $), $K^-$ 
($\ast$) and $\Delta$ ($\star$) in the central 125 fm$^3$ cell of 
Pb+Pb collisions at 160 AGeV at $t$=10 fm/$c$ are compared
to the UrQMD box calculations (open symbols) and 
to Boltzmann distributions (Eq.~\ref{spectra}) 
(lines) with  parameters $T$=161~MeV, 
$\mu_{\B}$=197~MeV, $\mu_{\rm S}=37$~MeV, 
obtained in the ideal gas SM.
}
\label{fig8}
\end{figure}

At $t>9$ fm/c the energy spectra of baryons $i$ in the  central
 cell of A+A collisions
are nicely reproduced
by Boltzmann  distributions
\begin{equation}
\frac{d^3N}{d^3p}~=~\frac{dN}{4\pi pE_idE_i}~=~
\frac{Vd_i}{(2\pi\hbar)^3} 
\exp\left(\frac{B_i\mu_B + S_i\mu_S}{T}\right)~
\exp\left(- \frac{E_i}{T}\right)
\label{spectra}
\end{equation}
 with parameters $T,\mu_B,\mu_S$
extracted by means of SM from Eqs.~(1-3) (see figure~\ref{fig8}). 
The meson UrQMD-spectra in the central cell at SPS energies 
%the UrQMD model 
exhibit systematically
% higher temperatures
% about 30 MeV 
lower inverse slope ("temperature") than $T$ found from SM.
It is not a result of collective motion which is negligible
inside the central cell. The same difference of baryon
and meson inverse slope "temperatures" are found for infinite matter 
UrQMD calculations \cite{Belk98}.
%gives
%then for baryons.

\begin{figure}  % Particle species in UrQMD cell and SM
\psfig{file=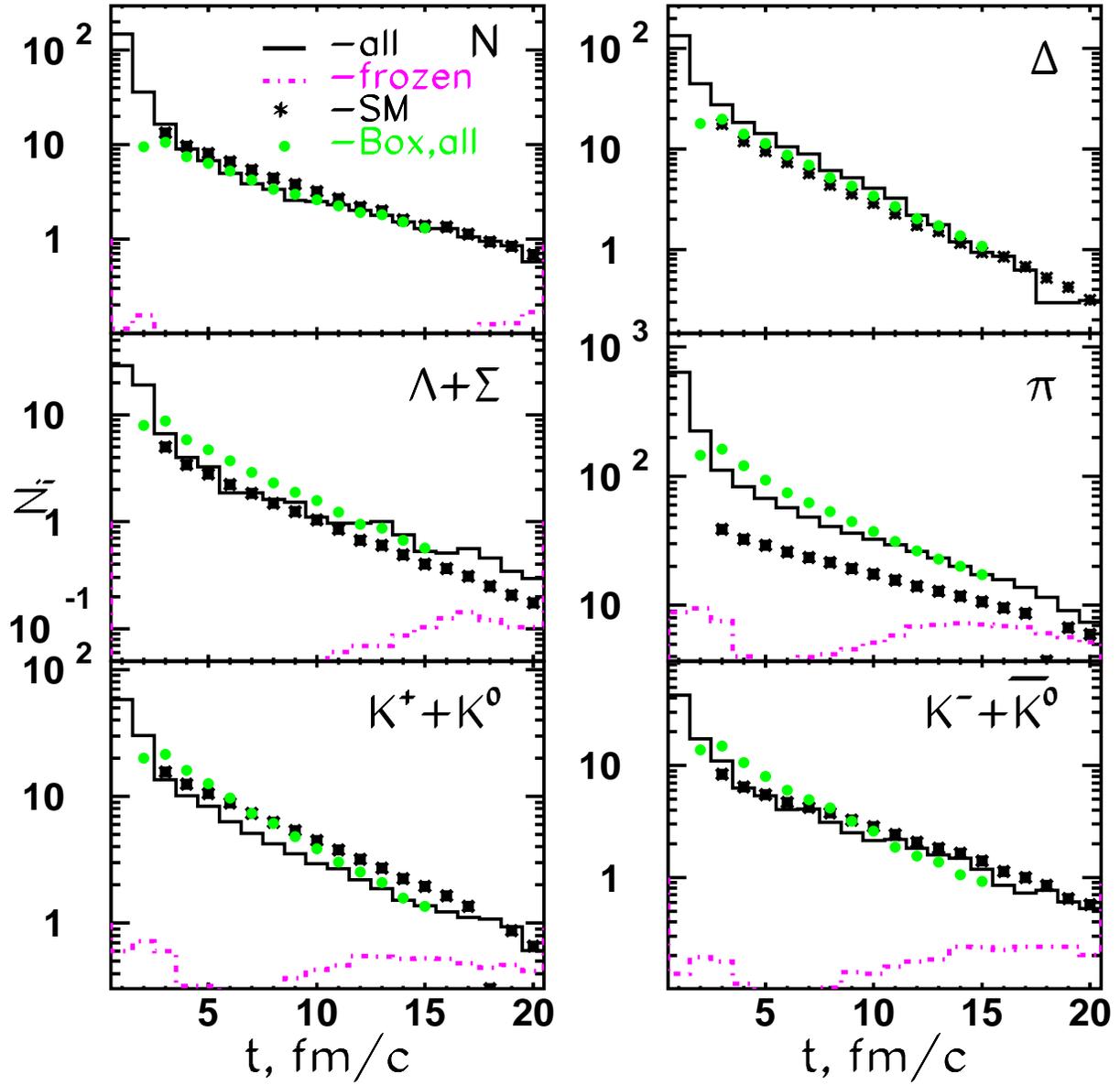,width=16cm}
\caption{ 
The number of particles in the central cell of Pb+Pb collisions at 
160 AGeV as a function of time as obtained in the UrQMD model 
(histograms) together with the predictions of the SM (stars) and
of the UrQMD box calculations (open symbols). Solid 
lines correspond to all hadrons in the cell, dot-dashed lines $-$ to 
the particles in the cell already frozen at time t. 
}
\label{fig9}
\end{figure}

A similar situation takes place for hadron yields: the  SM 
%(Eq.~(4)) 
nicely describes baryon and some meson production 
but significantly underestimates 
the pion yields obtained from UrQMD
in  heavy ion collisions  
at SPS energies
(see figure~\ref{fig9}). 
%The agreement is surprisingly good except of pions, which number 
%the SM underestimates by a factor of two. 
Note that the description of the pion yield is a problem 
for the ideal gas SM.
% has
%with the final state multiplicities. 
To get the agreement with the 
experimental data one needs to implement  some mechanisms
to enhance the pion multiplicity, e.g. effective chemical
potential for pions \cite{nucth98}. 

To make a final conclusion about local thermal and chemical equilibrium
in the central cell of relativistic heavy ion collisions 
simulated in  microscopic UrQMD one should make a detailed comparison with
the equilibrated infinite matter 
created within the same UrQMD model 
with the same parameters $(\epsilon, \rho_B, \rho_S)$ 
as in the cell at a certain time $t$. 
Here by words "equilibrated" we mean the hadronic matter in the box 
which in a course of many collisions 
reaches a) the saturation of different hadronic yields (chemical equilibrium)
and b) whose
spectra are not changed any more with further collisions 
(thermal equilibrium). The comparison between cell and box calculations
shows 
that at $t>9$ fm/c of Pb+Pb collisions at SPS 
energy spectra and hadron yields in the cell are in a very good agreement 
with box calculations (see figures~\ref{fig8},\ref{fig9}).
%Figure~\ref{fig10} shows the UrQMD
%simulations in the cell and in the box together with the SM 
%estimations. At the high density phase of a heavy ion collision the  
%cell results exhibits dramatic deviations from the equilibrium. But
%after $t \geq 10$ fm/$c$ there is no noticeable difference between 
%box calculations and the UrQMD cell with $V=5 \times 5 \times 5$ 
%in Pb+Pb collisions at 160 AGeV. There is a difference between
%the temperature, predicted by SM, and obtained 
%by Boltzmann fit of the particle spectra in UrQMD Pb+Pb calculations
%at 160 AGeV, especially 
Both UrQMD cell and box calculations show strong deviation from
SM results
at high energy densities. 
%As it was shown
%in \cite{Belk98}  
%The main reason of this deviation is the 
%large number of string degrees of freedom at high energies 
%which have not been included in our present version
%of SM equations (1-3).
It should be stressed that the UrQMD equilibrium has a principal difference
from SM approach. It corresponds to strong pion (or rather meson) 
yield enhancement at high energy density 
(formally, to $\mu_{\pi}>0$), appeared  
at initial non-equilibrium stage of the reaction. 
It is connected to
the absence of the reverse reactions to the multiple string decays 
$ string \rightarrow h_1+...+h_n$ with $n\ge 3$ and 
more than 2-body resonance decays $R \rightarrow h_1+h_2+...$
% , for $N \ge 3$ 
in UrQMD. 
% model. 
Note that at AGS collision energy the maximal energy density
%during the LTE stage 
is not too high, the produced hadronic matter is baryonic reach, and 
the agreement between the SM and UrQMD cell calculations for all 
hadronic species at $t\ge 10$ fm/c is satisfactory \cite{bravina98}. 

\begin{figure}  % EOS: $T(\mu)$
\psfig{file=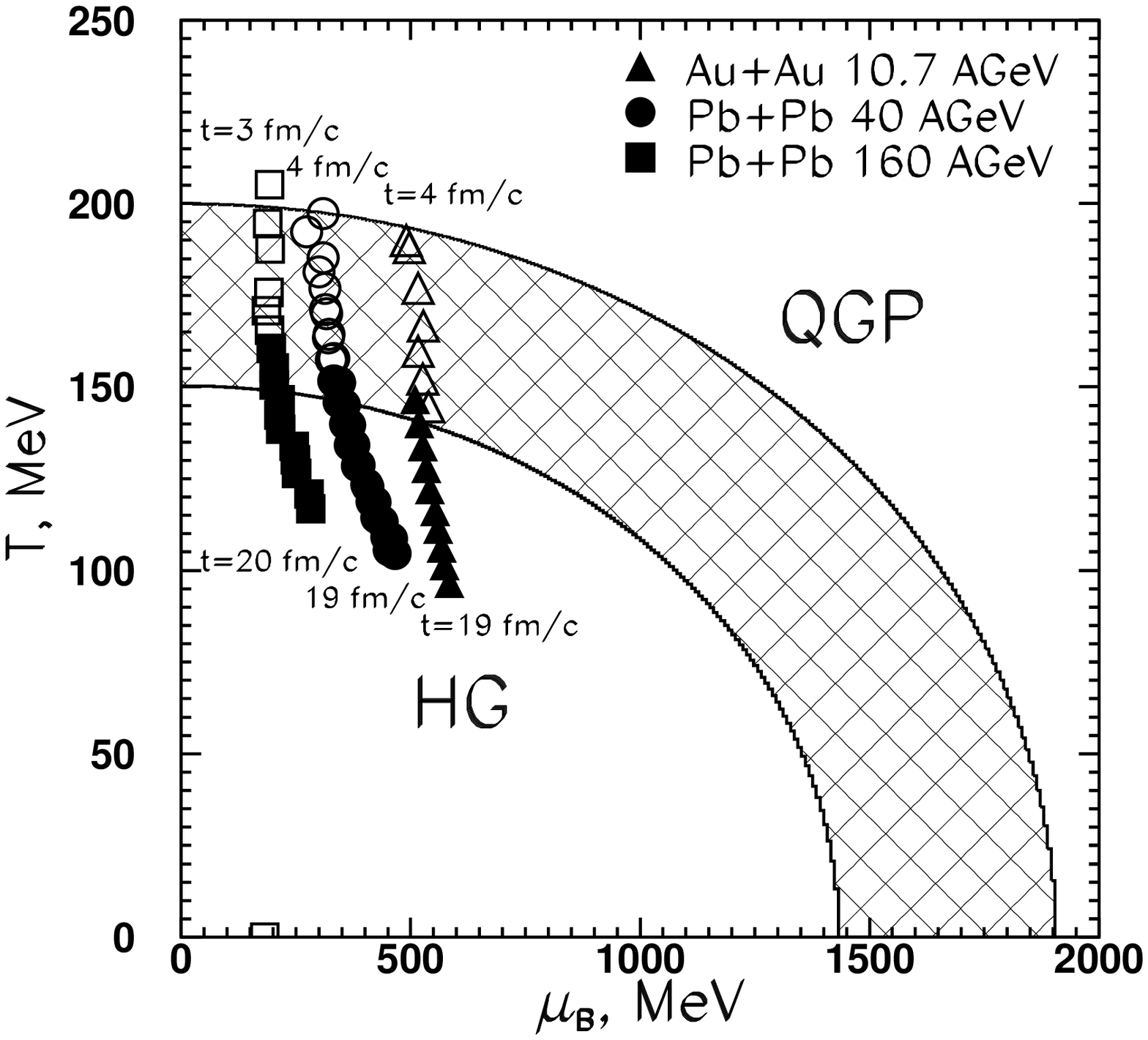,width=16cm}
\caption{
Temperature versus baryonic chemical potential
predicted by the SM for the time 
evolution of the hadronic characteristics $\{\varepsilon$, $\rho_B$,
$\rho_S\}$ obtained within the UrQMD in the central cell of central 
A+A collisions at 10.7, 40 and 160 AGeV.
(Open) full  symbols show the (non-) equilibrium stage of the reaction.
The two solid lines describe the boundary 
%  of the nonstrange
%free 
of the QGP  
calculated for two different bag constants, $B^{1/4}=227$ and 302 MeV
corresponding to $T_{\rm c}$=150 and 200 MeV at $\mu_{\B}=0$.
}
\label{fig11}
\end{figure}

Figure~\ref{fig11} shows the time evolution of the temperature, T, 
versus the baryonic chemical potential, $\mu_{\B}$. We see that 
average $\mu_{\B}$ in the reaction drops drastically with the initial collision energy, 
while the maximal temperature is growing and practically reaches the upper phase 
transition boundary with the critical temperature $T_c$=200 MeV, 
predicted by the MIT bag model (details of the used bag model 
see in \cite{goren97}). 

\section{Conclusions}
Equilibration of hadronic matter produced in the central zone of 
heavy ion collisions at the projectile energies from 10.7 to 160
AGeV has been studied in the microscopic UrQMD model. The details
about Au+Au collisions at AGS energy can be found in \cite{bravina98}.
For SPS energies the
following conclusions may be drawn. 

1. For time $t=2$ fm/$c$ the velocity distributions of nucleons 
become isotropic in the $4 \times 4 \times 1$ fm$^3$ cell.
Pions are equilibrated much later, at $t \cong 8$ fm/$c$. This
effect is caused by the non-zero formation time for non-leading
particles.

2. Temperature of particles in the cell coincides with the 
corresponding temperature in the box at $t \cong 10$ fm/$c$.

3. Spectra of all particle species in the cell start 
to be described by the 
UrQMD box calculations at $t \cong 10$ fm/$c$. Therefore, the
(local) thermal and chemical equilibrium in the central cell is
established at $t \sim 10$ fm/$c$, not earlier.

4. Using the statistical model of ideal gas one obtains the 
macroscopic characteristics, like pressure, temperature, chemical
potential, etc. of the system. The entropy per baryon is nearly 
constant, $s / \rho_B = 30 (15)$ at SPS (AGS) energies during the time 
interval $t = 10 - 18$ fm/$c$.

5. The equation of state obtained from the microscopic calculations 
(Eq.~\ref{pressure_micr}) 
as well as from the SM predictions (Eq.~\ref{pressure_sm}) 
in a whole energy range has a simple form $P \cong 
(0.12-0.15)\, \varepsilon$, which is 
%very 
close to that suggested many years ago for the
%by the statistical model of ideal 
resonance gas \cite{Shur72}.
%This analysis is encouraging but clearly awaits better data for a
%conclusive test of the local thermal equilibrium in heavy ion 
%collisions.

6. At high energy densities a deviation of the UrQMD 
equilibrium state from SM results is found. It is connected
with a specific behaviour of string degrees of freedom in the UrQMD model.
This energy density region lies close (or even above) to
the expected phase transition boundary of the quark-gluon plasma state.  
The study of this energy density region deserves further efforts.

\ack 
Fruitful discussions with L Csernai,  U Heinz, I Mishustin, B  M{\"u}ller, 
L Neise, J  Rafelsky,  J  Stachel and N Xu are 
greatfully acknowledged.
This work was supported by A v Humboldt Stiftung, GSI, DFG, BMBF, 
Graduiertenkolleg ``Exp u Theoret Schwerionenphysik'' and
Buchmann Stiftung.

\Figures

%\newpage

%\begin{figure}  % Directed flow
%\caption{ 
%Bounce-off flow $<p_x(y)/A>$-distribution of the final state hadrons
%vs. rapidity $y_{\rm cm}$ in Pb+Pb collisions (b=3 fm) at 160 
%AGeV. Due to initial momentum transfer the leading baryons ($N$, 
%$\Lambda + \Sigma$) have positive directed flow while all other 
%particles ($\bar{\Lambda}+\bar{\Sigma}$, $\pi$, $K$, $\bar{K}$) 
%have a small anti-flow effect.
%}
%\label{fig2}
%\end{figure}

\newpage

%\begin{figure}  %UrQMD box calculations
%\caption{ 
%Pressure vs. energy density calculated microscopically in
%the UrQMD for equilibrated infinite relativistic hadronic matter at  
%$\rho_{\rm S}=0$ and $\rho_{\B}=$1, 3 and 8$\rho_0$ (bold solid 
%line), and in central zones of A+A collisions at 10.7, 40 and 160
%AGeV at different times.
%}
%\label{fig4}
%\end{figure}

\newpage

\newpage

\newpage

\newpage

%\begin{figure}  % EOS T versus $\varepsilon$ - real fit of spectra.
%\caption{ 
%Time evolution of energy density and temperature in the central cell 
%of central (b=0) Pb+Pb collisions at 160 AGeV (red points) compared 
%to the equilibrated infinite hadron matter (black points) and to the 
%predictions of the ideal gas SM (blue solid curve) calculated with the 
%same $\varepsilon$, $\rho_{\B}$ and $\rho_{\rm S}$. Temperature was 
%extracted as an inverse slope parameter from the fit of the energy 
%$d^3N/d^3p$ spectra of $\Delta$, $N$, $\Lambda$, $\pi$, $K$ and 
%$\bar{K}$ to Boltzmann exponential distribution.
%}
%\label{fig10}
%\end{figure}

\newpage


\section*{References}
\begin{thebibliography}{99}

\bibitem{Landau}
Landau L D, 1953 {\it Izv. Akad. Nauk SSSR} {\bf 17} 51
\nonum Belenkij S Z and Landau L D, 1956 {\it Suppl.} \NC
{\bf 3} 15

\bibitem{Twofl} 
Amsden A A, Goldhaber A S, Harlow F H, Nix J R, 1978 \PR C {\bf 17} 
2080
\nonum Csernai L P {\it et al}, 1982 \PR C {\bf 26} 149
\nonum Clare R B, Strottman D, 1986 {\it Phys. Rep.} {\bf 141} 177
\nonum Barz H W, K\"ampfer B, Csernai L P, Lukacs B, 1987 
     \NP A {\bf 465} 743
%\nonum Barz H W, K\"ampfer B, 1988 \PL B {\bf 206} 399

\bibitem{BMGR}  Bernard S, Maruhn J A, Greiner W, Rischke D H,
1996 \NP A {\bf 605} 566

\bibitem{Strott} Schlei B R, Ornik U, Plumer M, Strottman D, 
Weiner R M, 1996 \PL B {\bf 376} 212
%\nonum Ornik U, Pottag F, Weiner R M, 1989 \PRL {\bf 63} 2641

\bibitem{hydro}
Brachmann J, Dumitru A, Maruhn J A, St\"ocker H, Greiner W, 
Rischke D H, 1997 \NP A {\bf 619} 391

\bibitem{Fermi} 
Fermi E, 1950 {\it Prog. Theor. Phys.} {\bf 5} 570;
1951 \PR {\bf 81} 683

\bibitem{Hagedorn} 
Hagedorn R, 1965 {\it Suppl.} \NC {\bf 3} 147;
\nonum
Hagedorn R and Rafelsky J, 1980 \PL B {\bf 97} 136

\bibitem{Braun}
Braun-Munzinger P, Stachel J, Wessel J P and Xu N,
1995 \PL B {\bf 344} 43; 1995 \PL B {\bf 365} 1 
%\nonum Stachel J, 1996 \NP A {\bf 610} 509c 

\bibitem{Goren} Yen G D, Gorenstein M I, Greiner W and Yang S N, 
1997 \PR C {\bf 56} 2210

\bibitem{Shur72}
Shuryak E, 1972 {\it Sov. J. Nucl. Phys.} {\bf 16} 395

\bibitem{Hofmann} 
Hofmann J, St\"ocker H, Heinz U, Scheid W, Greiner W, 
1976 \PRL {\bf 36} 88

\bibitem{Geiger}
Geiger K, 1995 {\it Phys. Rep.} {\bf 258} 237

\bibitem{Heinz}
Heinz U, \NP A {\it submitted}; nucl-th/9801050
\nonum Schnedermann E, Sollfrank J and Heinz U, 1993 \PR C {\bf 48} 
2462

\bibitem{urqmd1} 
Bass S A {\it et al}, 1998 {\it Prog. Part. Nucl. Phys.} {\bf 41}
225

\bibitem{QGSM}
Amelin N S, Bravina L V, Sarycheva L I, Smirnova L N, 1990 
{\it Sov. J. Nucl. Phys.} {\bf 51} 1093
\nonum Amelin N S, Bravina L V, Csernai L P, Toneev V D, Gudima K K
and Sivoklokov S Yu, 1993 \PR C {\bf 47} 2299

\bibitem{RQMD}
Sorge H, St{\"o}cker H and Greiner W, 1989 \APNY {\bf 192} 266

\bibitem{ARC}
Pang Y, Schagel T J and Kahana S H, 1992 \PRL {\bf 68} 2743

\bibitem{ART}
Bao-An Li and Che Ming Ko, 1995 \PR C {\bf 52} 2037; 
1996 \PR C {\bf 53} 22

\bibitem{Greiner}
St{\"o}cker H and Greiner W, 1986 {\it Phys. Rep.} {\bf 137} 277

\bibitem{bravina95} Bravina L V, Mishustin I N, Amelin N S,
Bondorf J, Csernai L P, 1995 \PL B {\bf 354} 196; 
1997 {\it Heavy Ion Phys.} {\bf 5} 455

\bibitem{Shur95} 
Hung C M and Shuryak E, 1995 \PRL {\bf 75} 4003; 
1997 \PR C {\bf 56} 453

\bibitem{Matiello9597}
Mattiello R, Jahns A, Sorge H, St{\"o}cker H and Greiner W,
1995 \PRL {\bf 74} 2180
\nonum Mattiello R, Sorge H, St{\"o}cker H and Greiner W,
1997 \PR C {\bf 55} 1443

\bibitem{Bass1}
Bass S A {\it et al}, \PRL {\it submitted}; nucl-th/9711032

\bibitem{Belk98} 
Belkacem M {\it et al}, 1998 \PR C {\bf 58} 1727

\bibitem{bravina98} 
Bravina L V {\it et al}, 1998 \PL B {\bf 434} 379

\bibitem{nucth98} 
Yen G D and Gorenstein M I, nucl-th/9808012

\bibitem{weber98}
H  Weber, C  Ernst, M  Bleicher, L  Bravina, H  Stoecker, 
W  Greiner, C   Spieles, S  A  Bass, 1998 nucl-th/9808021

\bibitem{berenguer}
M  Berenguer, C  Hartnack, G  Peilert, H  S\"ocker,
W  Greiner, J  Aichelin and A  Rosenhauer,
1992 {\it J. Phys.  G}  {\bf 18}   655 

\bibitem{sorge}
H  Sorge, 1997 \PRL {\bf 78} 2309;
1997 \PL B {\bf 402} 251 

\bibitem{goren97} 
Gorenstein M I {\it et al}, nucl-th/9711055.
%\bibitem{bravina98a} 
%Bravina L V {\it et al}, in preparation 

\end{thebibliography}
\end{document}